\documentclass[aps,pra,twocolumn,eqsecnum]{revtex4-2}
\usepackage{graphicx,color}
\usepackage{amsmath}
\usepackage{amssymb}
\usepackage{amsthm}
\usepackage{multirow}
\usepackage{hyperref}
\usepackage{enumitem}
\usepackage{bm}

\newtheorem{thm}{Theorem}
\newtheorem{prop}[thm]{Proposition}
\newtheorem{lemma}[thm]{Lemma}
\newtheorem{corol}[thm]{Corollary}

\theoremstyle{definition}

\newcommand{\rmd}{\mathrm{d}}
\newcommand{\rmi}{\mathrm{i}}

\newcommand{\Tr}{\operatorname{Tr}}

\newcommand{\T}{\operatorname{T}}

\definecolor{dgreen}{rgb}{0,0.5,0}

\definecolor{dblue}{rgb}{0,0,0.6}

\definecolor{dred}{rgb}{0.784,0,0}

\definecolor{dorange}{cmyk}{0,0.72,1,0.16}
\definecolor{dmagenta}{rgb}{0.847,0.149,0.490}

\definecolor{delete}{cmyk}{0.5,0,0,0}

\begin{document}
\title{No-Go Theorem for Quantum Heat Engines Powered Purely by Quantum Measurements in the Steady Regime}
\author{Kenta Koshihara}
\affiliation{Department of Physics, Waseda University, Tokyo 169-8555, Japan}
\author{Kazuya Yuasa}
\affiliation{Department of Physics, Waseda University, Tokyo 169-8555, Japan}
\date{\today}
\begin{abstract}
We study the thermodynamics of a quantum measurement-powered engine that converts energy injected by measurement backaction into work.
We consider an engine with a finite-dimensional working substance driven \textit{purely} by quantum measurements, i.e., by bare quantum measurements, without feedback control or thermal contact in the thermodynamic cycle.
On the basis of a Poincar\'e-like recurrence theorem for general quantum channels, we prove a no-go result for work extraction from such an engine in the steady regime.
In the steady regime, quantum measurements become nondisturbing and do not inject energy into the working substance.
Consequently, no work can be extracted.
This result reveals the necessity of an entropy-decreasing process, such as feedback control or thermal contact, for work extraction in steady-cycle measurement-powered engines.
\end{abstract}
\maketitle

%%%%%%%%%%%%%%%%%%%%%%%%%%%%%%%%%%%%%%%%%%%%%%%%%%%%%%%%%%%%%%%%%%%%%%%%%%%%%%%
\section{Introduction}
\label{section:Introduction}
A heat engine is a fundamental concept in thermodynamics.
In a typical setup, the second law of thermodynamics governs how much work can be extracted from an engine operating between thermal reservoirs at different temperatures.
Recent studies on Maxwell's demon have revealed that microscopic information gained through measurements can also be converted into work via feedback control~\cite{MaruyamaNoriVedral2009, ParrondoHorowitzSagawa2015, GooldHuberRiera2016}. 
In particular, it is possible to extract work even from a single thermal reservoir by performing feedback control~\cite{AbreuSeifert2011}. 
Generalized second laws that incorporate information as a physical resource have clarified the role of acquired information in scenarios where the working substance is fully thermalized through contact with thermal reservoirs~\cite{SagawaUeda2008, MorikuniTasaki2011, FunoWatanabeUeda2013}, as well as in scenarios where thermodynamic cycles are repeated without full thermalization and the engine operates in the steady regime~\cite{StrasbergSchallerBrandes2017, KoshiharaYuasa2022, FuWangChen2023}.

In the quantum regime, measurements not only acquire information but also induce backaction on the measured system.
In particular, when measuring superpositions of different energy eigenstates, the measurement process inevitably injects or dissipates energy. 
This feature can be exploited as a thermodynamic operation, leading to the concept of a \textit{quantum measurement-powered engine}~\cite{ElouardHerrera-MartiClusel2017}.
The energy injected by measurement backaction can be converted into work by applying feedback control~\cite{ElouardHerrera-MartiHuard2017, MohammadyAnders2017, ElouardJordan2018, BresqueCamatiRogers2021, ManikandanElouardMurch2022, WangXiaBresque2022, Rodin2023} and/or by bringing the working substance into contact with thermal reservoirs~\cite{YiTalknerKim2017, DingYiKim2018, DasGhosh2019, SuLinChen2021, AnkaDeOliveiraJonathan2021, LinSuChen2021, AlamVenkatesh2022, PurkaitBiswas2023, BhandariCzupryniakErdman2023, JussiauBresqueAuffeves2023, ElMakouriSlaouiAhlLaamara2023}. 
The role of generalized quantum measurements in measurement-powered engines has also been investigated~\cite{Behzadi2020, LisboaDieguezGuimaraes2022, DieguezLisboaSerra2023}.

\begin{figure*}
\includegraphics[scale=0.55]{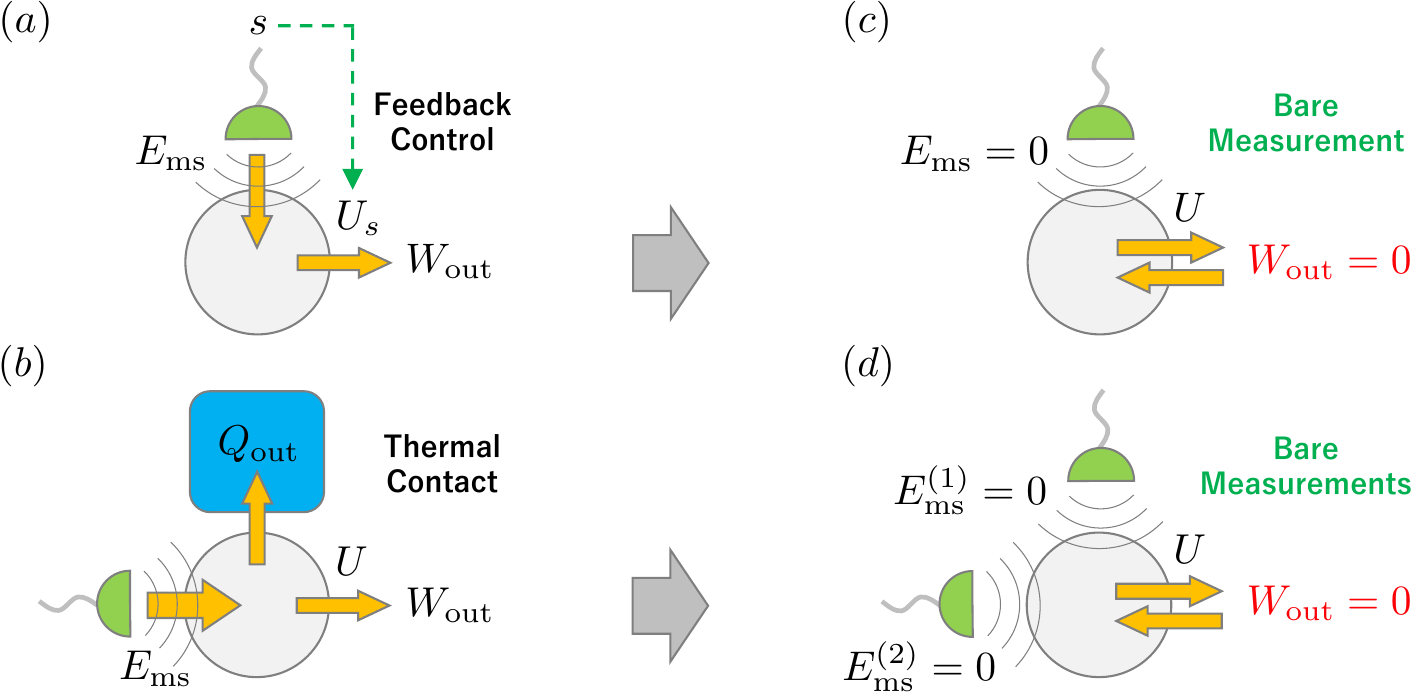}
\caption{
Quantum measurement-powered heat engines.
(a) Feedback-assisted measurement-powered engine~\cite{ElouardHerrera-MartiHuard2017}.
A quantum measurement is performed, yielding an outcome $s$, followed by a unitary feedback operation $U_s$ conditioned on $s$, through which work $W_\mathrm{out}$ is extracted.
(b) Thermalization-assisted measurement-powered engine~\cite{YiTalknerKim2017}.
A quantum measurement is performed, followed by a unitary operation $U$, independent of the measurement outcome, through which work $W_\mathrm{out}$ is extracted, and the working substance is brought into contact with a thermal reservoir.
(c) If the measurement outcome in (a) is not used for feedback, no work can be extracted ($W_\mathrm{out}=0$) in the steady cycle.
(d) If the thermal reservoir in (b) is replaced by a bare quantum measurement, no work can be extracted ($W_\mathrm{out}=0$) by a unitary operation in the steady cycle.
Generalizing this situation, we consider cycles involving multiple quantum measurements and multiple unitary operations in the present work (see Fig.~\ref{fig:PureMPE}).
}
\label{fig:MPEList}
\end{figure*}

More specifically, there are two main types of quantum measurement-powered engines.
One is ``thermodynamics without a heat bath''~\cite{ElouardHerrera-MartiHuard2017} [see Fig.~\ref{fig:MPEList}(a)]. 
In this type of engine, the energy injected by measurement backaction is directly converted into work via feedback control.
We refer to this type as \textit{feedback-assisted} measurement-powered engines. 
The other is the ``single-temperature engine without feedback control''~\cite{YiTalknerKim2017} [see Fig.~\ref{fig:MPEList}(b)].  
In this case, a quantum measurement injects energy into a working substance thermalized with a heat bath, and work is extracted through unitary control without utilizing the information acquired by the measurement.
In this scenario, the quantum measurement plays the role of a ``hot'' bath~\cite{footnote:MPEClausius}. %\footnote{This engine works even with a single reservoir, which effectively acts as a cold bath that absorbs heat. Since no feedback control is applied, the standard Clausius inequality $\beta Q \leq 0$ holds for the heat $Q$ and the inverse temperature $\beta$ of the reservoir. See Ref.~\cite{KoshiharaYuasa2022}.}. %
We refer to this type as \textit{thermalization-assisted} measurement-powered engines.

A natural question is whether work can be extracted from a quantum measurement-powered engine without feedback control or thermal contact.
The necessity of feedback control or thermal contact for such engines to work is commented in Ref.~\cite{JordanElouardAuffeves2020}, but a rigorous proof has been lacking.
In this paper, we provide such a proof in the case of a finite-dimensional working substance.

We consider a quantum engine driven \textit{purely} by quantum measurements, without feedback control or thermal contact [see Figs.~\ref{fig:MPEList}(c) and~(d)]. 
The key idea is to identify the genuine backaction of a quantum measurement.
It is important to note that a general quantum measurement is operationally indistinguishable from a measurement followed by feedback control.
In contrast, a \textit{bare} quantum measurement~\cite{Wiseman1995, FuchsJacobs2001} can be regarded as a \textit{pure} measurement without feedback.
See, e.g., Refs.~\cite{Wiseman1995, FuchsJacobs2001, Jacobs2009, KoshiharaYuasa2022} and Sec.~\ref{section:Setup}\@.
We prove that the backaction of a \textit{bare} quantum measurement inevitably increases the entropy of the measured system.
This leads to a no-go result for work extraction from an engine driven by bare quantum measurements operating in the steady regime.
In such a regime, the recurrence property of general quantum channels~\cite{Wolf2012, AmatoFacchiKonderak2023} implies that the entropy change over a cycle must vanish, and therefore each measurement must preserve the entropy.
As a result, the measurement cannot disturb the working substance, no energy can be injected by the measurement, and no work can be extracted from the engine.
This establishes the necessity of an entropy-decreasing process, such as feedback control or thermalization, for a quantum measurement-powered engine to work in the steady regime.

One of the characteristics of bare quantum measurements is that they are minimally disturbing~\cite{Wiseman1995, FuchsJacobs2001, Banaszek2001, Barnum2002}.
Another important property is their unitality, which has been found to be relevant in quantum thermodynamics.
In particular, unitality is relevant for fluctuation theorems~\cite{MorikuniTasaki2011, Rastegin2013, GooldModi2021}, is essential for passivity~\cite{AnkaDeOliveiraJonathan2021} and the ergotropy bound~\cite{SolfanelliBuffoniCuccoli2019, KoshiharaYuasa2023}, and is consistent with the standard Clausius inequality in the absence of feedback control~\cite{KoshiharaYuasa2022, BuffoniSolfanelliVerrucchi2019}.
These observations suggest that unitality captures fundamental thermodynamic constraints associated with quantum measurements.
The present work contributes to this perspective by showing that bare quantum measurements (a class of unital quantum operations) cannot by themselves power a heat engine in the steady regime, thereby establishing a limitation on measurement-induced energetics.

The paper is organized as follows.
In Sec.~\ref{section:PureMPE}, we introduce the thermodynamic cycle studied in this work, namely a heat engine powered \textit{purely} by quantum measurements, involving neither feedback control nor thermal contact.
In Sec.~\ref{section:SteadyCycle}, we point out that, as the cycle is repeated, the engine asymptotically enters the steady regime, in which work extraction is analyzed.
In Sec.~\ref{section:1stLaw}, we formulate the first law of thermodynamics for the steady cycle, and in Sec.~\ref{section:NoGoResult}, we prove a no-go result for work extraction in this regime.
This result implies the necessity of an entropy-decreasing process for work extraction.
In Sec.~\ref{section:Discussion}, we discuss why feedback-assisted~\cite{ElouardHerrera-MartiHuard2017} and thermalization-assisted~\cite{YiTalknerKim2017} measurement-powered engines can extract work.
Finally, we conclude in Sec.~\ref{section:Conclusion}\@.
Technical proofs essential to the main results are presented in Appendices~\ref{appendix:ProofMainTheorem}--\ref{app:AlternativeProof}\@.

\begin{figure*}
\includegraphics[width=0.95\textwidth]{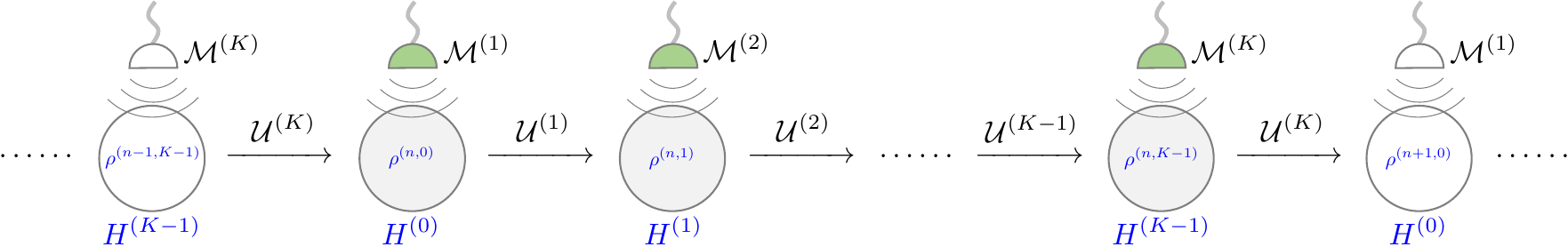}
\caption{
The cycle~(\ref{eq:TheCycle}) of a quantum engine powered \textit{purely} by quantum measurements.
In each cycle, $K$ pairs of nonselective bare quantum measurements $\mathcal{M}^{(k)}$ and unitary operations $\mathcal{U}^{(k)}$ are performed, where each unitary $\mathcal{U}^{(k)}$ is generated by driving the Hamiltonian as $H^{(k-1)} \to H^{(k)}$ ($k=1,\ldots,K$).
In the $n$th cycle, the state of the working substance evolves as $\rho^{(n,k)}=(\mathcal{U}^{(k)}\circ\mathcal{M}^{(k)})(\rho^{(n,k-1)})$, with $\rho^{(n,K)}=\mathcal{E}(\rho^{(n,0)})=\rho^{(n+1,0)}$.
To close the cycle, the Hamiltonian is returned to its initial one, $H^{(K)}=H^{(0)}$, at the end of each cycle.
In contrast to previously studied measurement-powered engines~\cite{ElouardHerrera-MartiHuard2017,YiTalknerKim2017}, this engine involves neither feedback control nor thermal contact.
Note that Fig.~\ref{fig:MPEList}(d) can be regarded as a special case of this general cycle with $K=2$, where $\mathcal{U}^{(1)}$ is generated by a cyclic driving of a Hamiltonian, so that the Hamiltonian returns to its initial form after $\mathcal{U}^{(1)}$, while the operation $\mathcal{U}^{(2)}$ following the second measurement is taken to be the identity map, $\mathcal{U}^{(2)}=\mathcal{I}$.
}
\label{fig:PureMPE}
\end{figure*}

%%%%%%%%%%%%%%%%%%%%%%%%%%%%%%%%%%%%%%%%%%%%%%%%%%%%%%%%%%%%%%%%%%%%%%%%%%%%%%%
\section{Heat Engine Powered Purely by Quantum Measurements}
\label{section:Setup}
%%%%%%%%%%%%%%%%%%%%%%%%%%%%%%%%%%%%%%%%%%%%%%%%%%%%%%%%%%%%%%%%%%%%%%%%%%%%%%%
\subsection{Cycle}
\label{section:PureMPE}
Let us consider a finite-dimensional quantum system with initial state $\rho^{(0)}$ and initial Hamiltonian $H^{(0)}$. 
Using this system as a working substance, we construct a quantum heat engine powered \textit{purely} by quantum measurements.
That is, the energy source of this engine is provided solely by the backaction of quantum measurements: no feedback control is applied and no thermal contact is introduced.
The cycle of this engine consists of the following steps (see Fig.~\ref{fig:PureMPE}).
\begin{enumerate}
\item 
We perform a measurement on the working substance to inject energy through the backaction of the measurement. 
We do not retain the outcome of the measurement. 
Such a nonselective quantum measurement is described by a completely positive and trace-preserving (CPTP) map $\rho^{(0)}\to\tilde{\rho}^{(1)}=\mathcal{M}^{(1)}(\rho^{(0)})$.

\item
We drive the Hamiltonian $H^{(0)}\to H^{(1)}$ by an external force to extract work from the engine. 
Under this control, the working substance evolves unitarily as
$\tilde{\rho}^{(1)}\to\rho^{(1)}=\mathcal{U}^{(1)}(\tilde{\rho}^{(1)})$.

\item
We repeat these operations $K$ times, with a fixed sequence of measurement and driving.
At the $k$th step, we perform a measurement $\mathcal{M}^{(k)}$ on $\rho^{(k-1)}$, followed by a driving $H^{(k-1)}\to H^{(k)}$ inducing a unitary $\mathcal{U}^{(k)}$.

\item
To close the cycle, we bring the Hamiltonian back to its initial one $H^{(K)}=H^{(0)}$ by the last $K$th driving.
\end{enumerate}
The evolution of the working substance per cycle is described by the CPTP map
\begin{equation}
\mathcal{E}
=
\mathcal{U}^{(K)}
\circ
\mathcal{M}^{(K)}
\circ
\cdots
\circ
\mathcal{U}^{(1)}
\circ
\mathcal{M}^{(1)}.
\label{eq:TheCycle}
\end{equation}
Repeated application of this cycle is described by $\mathcal{E}^n$ with $n=1,2,\ldots$\@

Here, the $k$th driving from $H^{(k-1)}$ to $H^{(k)}$ is described by a time-dependent Hamiltonian $H_k(t)$, with $H_k(0)=H^{(k-1)}$ and $H_k(\tau_k)=H^{(k)}$ at the beginning and end of the $k$th driving, respectively.
The unitary operation $\mathcal{U}^{(k)}(\rho)=U^{(k)}\rho U^{(k)\dag}$ is then generated by the Hamiltonian $H_k(t)$ as $U^{(k)}=\T\exp[-\rmi\int_0^{\tau_k}\rmd t\,H_k(t)]$.

In addition, we restrict $\mathcal{M}^{(k)}$ ($k=1,\ldots,K$) to \textit{bare} quantum measurements.
Nonselective bare quantum measurements are described by CPTP maps 
\begin{equation}
\mathcal{M}^{(k)}(\rho)=\sum_sM_s^{(k)}\rho M_s^{(k)},
\quad
M_s^{(k)}\geq0,
\label{eq:BareMeasurement}
\end{equation}
with positive-semidefinite measurement operators $M_s^{(k)}$ satisfying the Kraus condition $\sum_s(M_s^{(k)})^2=\openone$~\cite{Jacobs2014}.
They are also called \textit{minimally disturbing} measurements~\cite{WisemanMilburn2009}. 
A L\"uders projective measurement $\mathcal{M}(\rho)=\sum_sP_s\rho P_s$, with orthogonal projection operators $P_s$ satisfying $P_s=P_s^\dagger$, $P_sP_{s'}=\delta_{ss'}P_s$, and $\sum_sP_s=\openone$, is a typical example of a bare quantum measurement, since $P_s\geq0$\@. 
Note that a more general bare quantum measurement can be realized through an indirect projective measurement based on the Naimark representation theorem~\cite{HayashiIshizakaKawachi2015,Watrous2018}.

The bare (minimally disturbing) quantum measurements $\mathcal{M}^{(k)}$ should be distinguished from general quantum measurements, which are described by
\begin{equation}
\mathcal{M}_\mathrm{general}(\rho)=\sum_sK_s\rho K_s^\dag,
\label{eq:GeneralMeasurement}
\end{equation}
with general non-Hermitian Kraus operators $K_s$ satisfying the Kraus condition $\sum_sK_s^\dag K_s=\openone$~\cite{Ozawa1984, WisemanMilburn2009, Jacobs2014, HayashiIshizakaKawachi2015}.
A general quantum measurement $\mathcal{M}_\mathrm{general}$ can be regarded as (or is operationally indistinguishable from) quantum feedback control.
Indeed, by the polar decomposition of each Kraus operator,
\begin{equation}
K_s=U_sM_s,\quad M_s\ge0,	
\end{equation}
with $U_s$ unitary and $M_s\ge0$ positive-semidefinite, the general quantum measurement $\mathcal{M}_\mathrm{general}$ can be written as
\begin{equation}
\mathcal{M}_\mathrm{general}(\rho)=\sum_sU_sM_s\rho M_sU_s^\dag.
\label{eq:FeedbackControl}
\end{equation}
This can be interpreted as a quantum feedback process in which a bare quantum measurement with measurement operators $M_s$ is first performed, followed by a unitary control $U_s$ applied depending on the outcome $s$ of the measurement.
Note that the positive-semidefinite operators $M_s$ satisfy the Kraus condition
$\sum_sM_s^2=\sum_sK_s^\dag K_s=\openone$, and therefore constitute valid measurement operators.

In the heat engine studied here, we perform bare quantum measurements $\mathcal{M}^{(k)}$, each followed by a unitary control $\mathcal{U}^{(k)}$.
The bare measurements $\mathcal{M}^{(k)}$ are genuine measurements without feedback control, and the unitary controls $\mathcal{U}^{(k)}$ are applied irrespective of the outcomes of the measurements $\mathcal{M}^{(k)}$.
This engine appears to be powered \textit{purely} by quantum measurements.
However, we will show that the bare quantum measurements do not supply energy to the engine in the steady-cycle regime, and that the engine is therefore not actually powered by the measurements.
To activate the engine, it is necessary to introduce entropy-decreasing processes, such as feedback control or thermal contact.

%%%%%%%%%%%%%%%%%%%%%%%%%%%%%%%%%%%%%%%%%%%%%%%%%%%%%%%%%%%%%%%%%%%%%%%%%%%%%%%
\subsection{Steady Regime}
\label{section:SteadyCycle}
As the cycle described in the previous subsection is repeated, the heat engine enters the steady regime.
Mathematically, the evolution under repeated application of the CPTP map $\mathcal{E}$ defined in~(\ref{eq:TheCycle}) is asymptotically governed by the \textit{peripheral subspace} of $\mathcal{E}$~\cite{Wolf2012, BurgarthFacchiNakazato2020, AmatoFacchiKonderak2023, Hahn2025}.
We study work extraction in this steady regime.

Since the spectrum of a general CPTP map $\mathcal{E}$ is confined in the unit disk on the complex plane~\cite{Wolf2012, BurgarthFacchiNakazato2020, AmatoFacchiKonderak2023, Hahn2025, footnote:2}, %\footnote{This is because every CPTP map is, as a consequence of the Russo-Dye theorem, contractive with respect to the trace norm~\cite{Paulsen2003, Bhatia2009, Hiai2010}.},
only the eigenspaces of $\mathcal{E}$ associated with eigenvalues of unit magnitude survive in the asymptotic evolution of $\mathcal{E}^n$~\cite{Wolf2012, BurgarthFacchiNakazato2020, AmatoFacchiKonderak2023, Hahn2025}. 
The subspace spanned by these eigenspaces is called the \textit{peripheral subspace}, since it corresponds to the \textit{peripheral eigenvalues} on the unit circle.
The asymptotic evolution is described by
\begin{equation}
\mathcal{E}^n(\rho^{(0)})\to\mathcal{E}^n(\rho_\varphi)\quad\mathrm{as}\quad n\to\infty,
\label{eq:SteadyRegime}
\end{equation}
where
\begin{equation}
\rho_\varphi
=
\mathcal{P}_\varphi(\rho^{(0)})
\label{eq:PeripheralPart}
\end{equation}
is the projection of an arbitrary initial state $\rho^{(0)}$ onto the peripheral subspace of $\mathcal{E}$. 
Note that the peripheral projection $\mathcal{P}_\varphi$ is CPTP~\cite{Wolf2012, BurgarthFacchiNakazato2020, AmatoFacchiKonderak2023, Hahn2025}. 
The steady regime is characterized by the projected initial state $\rho_\varphi$.

Recall that, by a Poincar\'{e}-like recurrence theorem, there exists an increasing subsequence $\{n_i\}\subset\mathbb{N}$ ($0<n_1<n_2<\cdots$) such that $\mathcal{E}^{n_i}(\rho_\varphi)$ returns arbitrarily close to the projected initial state $\rho_\varphi$ as $i\to\infty$~\cite{Wolf2012, AmatoFacchiKonderak2023}, i.e.,
\begin{equation}
\mathcal{E}^{n_i}(\rho_\varphi)
\to
\rho_\varphi\quad\mathrm{as}\quad i\to\infty.
\label{eq:Recurrence}
\end{equation}
This includes various types of steady cycles, such as a fixed-point cycle $\mathcal{E}(\rho_\varphi)=\rho_\varphi$ starting from a fixed point $\rho_\varphi$ of the map $\mathcal{E}$, and a limit cycle satisfying $\mathcal{E}^m(\rho_\varphi)=\rho_\varphi$ for some $m\in\mathbb{N}$.
We are going to study the thermodynamics of the quantum measurement-powered engine in the general steady cycle.
In the following analysis, only the recurrence property~(\ref{eq:Recurrence}) is relevant, and the main results of this work hold for any steady regime arising from an arbitrary choice of the initial state $\rho^{(0)}$ of the engine. 
Note that, since the working substance is finite-dimensional in this work, all norms on its operator space are equivalent, and the limits in~(\ref{eq:SteadyRegime}) and~(\ref{eq:Recurrence}) hold with respect to any norm.

%%%%%%%%%%%%%%%%%%%%%%%%%%%%%%%%%%%%%%%%%%%%%%%%%%%%%%%%%%%%%%%%%%%%%%%%%%%%%%%
\section{Thermodynamics of Quantum Measurement-Powered Engine}
\label{section:Thermodynamics}
%%%%%%%%%%%%%%%%%%%%%%%%%%%%%%%%%%%%%%%%%%%%%%%%%%%%%%%%%%%%%%%%%%%%%%%%%%%%%%%
\subsection{First Law of Thermodynamics}
\label{section:1stLaw}
As the cycle is repeated, the evolution of the working substance enters the steady regime described by~(\ref{eq:SteadyRegime}). 
Let us consider the cycles $\mathcal{E}^n(\rho_\varphi)$ from the projected initial state $\rho_\varphi$.
Let $\rho_\varphi^{(n,k)}$ denote the state of the working substance after the $k$th step in the $n$th cycle, where $n=1,2,\ldots$ and $k=1,\ldots,K$.
These states are given recursively by
\begin{equation}
\rho_\varphi^{(n,k)}
=
(\mathcal{U}^{(k)}\circ\mathcal{M}^{(k)})(\rho_\varphi^{(n,k-1)}),
\label{eq:SteadyCycleStates}
\end{equation}
with $\rho_\varphi^{(1,0)}=\rho_\varphi$ and $\rho_\varphi^{(n+1,0)}=\rho_\varphi^{(n,K)}$.
See Fig.~\ref{fig:PureMPE}\@.

We define work as the energy change induced in the working substance by the unitary operations $\mathcal{U}^{(k)}$~\cite{PuszWoronowicz1978,Lenard1978,Alicki1979,AndersGiovannetti2013}. 
The work done on the working substance by the $k$th unitary operation $\mathcal{U}^{(k)}$ in the $n$th cycle
is given by~\cite{footnote:UnitaryWorkDerivation}
\begin{align}
W_U^{(n,k)}=
\Tr[H^{(k)}(\mathcal{U}^{(k)}\circ\mathcal{M}^{(k)})(\rho_\varphi^{(n,k-1)})]
&
\nonumber\\
{}-\Tr[H^{(k-1)}\mathcal{M}^{(k)}(\rho_{\varphi}^{(n,k-1)})]
&.
\label{eq:UnitaryWork}
\end{align}
Here, the initial and final Hamiltonians of a cycle coincide, i.e., $H^{(K)}=H^{(0)}$.
On the other hand, the energy injected by the $k$th bare quantum measurement $\mathcal{M}^{(k)}$ in the $n$th cycle is given by
\begin{align}
E_\mathrm{ms}^{(n,k)}
=
\Tr[H^{(k-1)}\mathcal{M}^{(k)}(\rho_{\varphi}^{(n,k-1)})]
&
\nonumber\\
{}
-\Tr[H^{(k-1)}\rho_\varphi^{(n,k-1)}]
&.
\label{eq:BareMeasurementEnergyGain}
\end{align}

Recall here that there exists an increasing subsequence $\{n_i\}\subset\mathbb{N}$ such that the working substance asymptotically returns to the projected initial state $\rho_\varphi$ as~(\ref{eq:Recurrence}), namely $\rho_\varphi^{(n_i,K)}\to\rho_\varphi$ as $i\to\infty$.
For such a subsequence $\{n_i\}$, the total energy change $(\Delta E)_{n_i}$ in the working substance after $n_i$ cycles is given by
\begin{align}
(\Delta E)_{n_i}
&=\sum_{n=1}^{n_i}\sum_{k=1}^K
(
W_U^{(n,k)}
+
E_\mathrm{ms}^{(n,k)}
)
\nonumber\\
&
=
\Tr[H^{(0)}\mathcal{E}^{n_i}(\rho_\varphi)]
-
\Tr[H^{(0)}\rho_\varphi]
\to0,
\end{align}
as $i\to\infty$.
By defining the net input work and the net energy injected by the bare measurements by
\begin{align}
W_U^\mathrm{total}&=\lim_{i\to\infty}\sum_{n=1}^{n_i}\sum_{k=1}^KW_U^{(n,k)},
\label{eq:WTotal}\\
E_\mathrm{ms}^\mathrm{total}&=\lim_{i\to\infty}\sum_{n=1}^{n_i}\sum_{k=1}^KE_\mathrm{ms}^{(n,k)},
\label{eq:EmsTotal}
\end{align}
we have
\begin{equation}
W_U^\mathrm{total}+E_\mathrm{ms}^\mathrm{total}=0.
\label{eq:EnergyConservation}
\end{equation}
This relation represents the first law of thermodynamics for the present engine in the steady regime.
The limits in~(\ref{eq:WTotal}) and~(\ref{eq:EmsTotal}) exist: we will see in the next subsection that $W_U^\mathrm{total}=E_\mathrm{ms}^\mathrm{total}=0$.

%%%%%%%%%%%%%%%%%%%%%%%%%%%%%%%%%%%%%%%%%%%%%%%%%%%%%%%%%%%%%%%%%%%%%%%%%%%%%%%
\subsection{No-Go Result for Work Extraction}
\label{section:NoGoResult}
In a measurement-powered engine, quantum measurements are expected to act as energy sources.
In the steady regime, however, this is not the case.
In this subsection, we prove that every bare quantum measurement $\mathcal{M}^{(k)}$ performed in the cycle does not supply energy to the working substance in the steady regime, that is,
\begin{equation}
E_\mathrm{ms}^{(n,k)}=0,
\quad
\forall n,k.
\label{eq:TheMainResult}
\end{equation}
This implies $E_\mathrm{ms}^\mathrm{total}=0$, and therefore,
\begin{equation}
W_U^\mathrm{total}=0,
\label{eq:NoWork}
\end{equation}
due to the first law of thermodynamics~(\ref{eq:EnergyConservation}): we can neither extract work from the engine nor inject energy into it in the steady regime.
This is the main result of this work.

The key idea is to quantify the backaction of the quantum measurement via the entropy gain.
The change in entropy due to the $k$th measurement $\mathcal{M}^{(k)}$ in the $n$th cycle is given by
\begin{equation}
\Delta S_\mathrm{ms}^{(n,k)}
=
S\bm{(}\mathcal{M}^{(k)}(\rho_{\varphi}^{(n,k-1)})\bm{)}
-
S(\rho_{\varphi}^{(n,k-1)}),
\label{eq:EntropyOfDisturbance}
\end{equation}
where $S(\rho)=-\Tr(\rho\log\rho)$ is the von Neumann entropy of the state $\rho$.
The quantity~(\ref{eq:EntropyOfDisturbance}) is called \textit{entropy of disturbance}~\cite{Jacobs2014}.
It quantifies the loss of information by the nonselective measurement $\mathcal{M}^{(k)}$~\cite{FuchsJacobs2001}.

A well-known property of the entropy of disturbance~(\ref{eq:EntropyOfDisturbance}) is monotonicity: any nonselective bare quantum measurement $\mathcal{M}(\rho)=\sum_sM_s\rho M_s$ with positive-semidefinite measurement operators $M_s$ does not decrease the entropy~\cite{Jacobs2014},
\begin{equation}
\Delta S_\mathrm{ms}
=S\bm{(}\mathcal{M}(\rho)\bm{)}-S(\rho)
\geq0.
\label{eq:UnitalEntropyGain}
\end{equation}
This follows from the unitality $\mathcal{M}(\openone)=\openone$ of the bare quantum measurement $\mathcal{M}$~\cite{footnote:UnitalMonotonicity}. %\footnote{In general, any unital CPTP map $\Phi$, satisfying the unitality condition $\Phi(\openone)=\openone$, does not decrease the von Neumann entropy $S(\rho)$. See e.g.~Refs.~\cite{AlickiFannes2001, Holevo2019, Sagawa2022}.}.
Moreover, we provide the necessary and sufficient condition for the equality to hold in~(\ref{eq:UnitalEntropyGain}),
\begin{equation}
\Delta S_\mathrm{ms}=0
\ %
\Leftrightarrow
\ %
\mathcal{M}(\rho)=\rho
.
\label{eq:MainTheorem}
\end{equation}
This is proved in Appendix~\ref{appendix:ProofMainTheorem}\@.

Now, take a subsequence $\{n_i\}$ leading to the recurrence~(\ref{eq:Recurrence}), and consider the overall entropy change $(\Delta S)_{n_i}$ in the working substance after $n_i$ cycles.
It vanishes as
\begin{equation}
(\Delta S)_{n_i}
=S\bm{(}\mathcal{E}^{n_i}(\rho_\varphi)\bm{)}-S(\rho_\varphi)\to0,
\end{equation}
as $i\to\infty$, due to the continuity of the von Neumann entropy~\cite{AlickiFannes2001, NielsenChuang2012, Wilde2017}.
Since the unitary operations $\mathcal{U}^{(k)}$ do not change the entropy, $\Delta S_U^{(n,k)}=S\bm{(}(\mathcal{U}^{(k)}\circ\mathcal{M}^{(k)})(\rho_\varphi^{(n,k-1)})\bm{)}-S\bm{(}\mathcal{M}^{(k)}(\rho_\varphi^{(n,k-1)})\bm{)}=0$, the overall entropy gain $(\Delta S)_{n_i}$ arises solely from the entropy of disturbance $\Delta S_\mathrm{ms}^{(n,k)}$.
We thus have
\begin{equation}
\Delta S^\mathrm{total}
=\lim_{i\to\infty}(\Delta S)_{n_i}
=\lim_{i\to\infty}\sum_{n=1}^{n_i}\sum_{k=1}^K\Delta S_\mathrm{ms}^{(n,k)}
=0.
\end{equation}
Since each $\Delta S_\mathrm{ms}^{(n,k)}$ is nonnegative, $\Delta S_\mathrm{ms}^{(n,k)}\ge0$, this implies
\begin{equation}
\Delta S_\mathrm{ms}^{(n,k)}=0
,
\quad
\forall n,k.
\label{eq:EntropyPreservingMeasurements}
\end{equation}
Moreover, due to the necessary and sufficient condition~(\ref{eq:MainTheorem}) for the equality, this further implies
\begin{equation}
\mathcal{M}^{(k)}(\rho_{\varphi}^{(n,k-1)})=\rho_{\varphi}^{(n,k-1)},
\quad
\forall n,k.
\label{eq:NondisturbanceSteady}
\end{equation}
That is, every bare quantum measurement $\mathcal{M}^{(k)}$ is nondisturbing in the steady regime, and therefore does not change the energy of the working substance.
Indeed, substituting~(\ref{eq:NondisturbanceSteady}) into the definition~(\ref{eq:BareMeasurementEnergyGain}) of the energy injected by the measurement, we obtain
\begin{align}
E_\mathrm{ms}^{(n,k)}
&=
\Tr[
H^{(k-1)}
\mathcal{M}^{(k)}(\rho_\varphi^{(n,k-1)})
]
-
\Tr[
H^{(k-1)}\rho_\varphi^{(n,k-1)}
]
\nonumber\\
&=0,
\quad \forall n,k,
\end{align}
which proves~(\ref{eq:TheMainResult}).
Consequently, $E_\mathrm{ms}^\mathrm{total}$ defined in~(\ref{eq:EmsTotal}) vanishes, $E_\mathrm{ms}^\mathrm{total}=0$\@.
The first law of thermodynamics in~(\ref{eq:EnergyConservation}) then implies $W_U^{\mathrm{total}}=0$\@.
This completes the proof of the no-go result~(\ref{eq:NoWork}).

%%%%%%%%%%%%%%%%%%%%%%%%%%%%%%%%%%%%%%%%%%%%%%%%%%%%%%%%%%%%%%%%%%%%%%%%%%%%%%%
\subsection{Necessity of Entropy-Decreasing Operation}
\label{section:Discussion}
The net entropy change vanishes, $\Delta S^\mathrm{total}=0$, in the steady regime, while bare quantum measurements $\mathcal{M}^{(k)}$ do not decrease the entropy.
This prevents bare quantum measurements from playing an active role in engines powered purely by quantum measurements.
In order to ``activate'' the backaction of quantum measurements, it is necessary to include an entropy-decreasing process in the cycle.
This explains why feedback-assisted measurement-powered engines~\cite{ElouardHerrera-MartiHuard2017} and thermalization-assisted measurement-powered engines~\cite{YiTalknerKim2017} can work.
Since feedback control and thermalization processes can decrease the entropy of the working substance, bare quantum measurements are allowed to increase the entropy, consistently with $\Delta S^\mathrm{total}=0$.
The entropy-decreasing mechanism balances the entropy increase induced by measurement backaction.

By feeding back the information acquired through a measurement, the entropy of the working substance can be reduced.
This mechanism is exploited in the quantum Maxwell demon~\cite{SagawaUeda2008}.
A quantum measurement followed by unitary feedback control is described by a CPTP map $\mathcal{E}_{\mathrm{ms}+\mathrm{fb}}(\rho)=\sum_s U_s M_s \rho M_s U_s^\dag$, where $M_s \ge 0$ are positive-semidefinite measurement operators satisfying the Kraus condition $\sum_s M_s^2=\openone$, and $U_s$ are unitaries applied depending on the outcome $s$ of the measurement.
The entropy change due to this operation is given by $\Delta S_{\mathrm{ms}+\mathrm{fb}}=\Delta S_\mathrm{ms}+\Delta S_\mathrm{fb}$, where the first contribution corresponds to the measurement, given by~(\ref{eq:UnitalEntropyGain}), and the second contribution
\begin{equation}
\Delta S_\mathrm{fb}=S\biggl(\sum_s U_s M_s \rho M_s U_s^\dag\biggr)-S\biggl(\sum_s M_s \rho M_s\biggr)
\end{equation}
corresponds to the unitary feedback control.
The latter is bounded as~\cite{SagawaUeda2008, KoshiharaYuasa2022}
\begin{equation}
\Delta S_\mathrm{fb}\ge -\chi_\mathrm{ms},
\label{eq:DeltaSfb}
\end{equation}
where
\begin{equation}
\chi_\mathrm{ms}=S\biggl(\sum_s M_s \rho M_s\biggr)-\sum_s p_s S(M_s \rho M_s/p_s),
\end{equation}
with $p_s=\Tr(M_s^2\rho)$, is the information gain associated with the measurement~\cite{footnote:InformationGain}. %\footnote{The quantity $\chi_\mathrm{ms}$ is the quantum mutual information (i.e.~correlation) between the measured system and a ``memory'' system~\cite{KoshiharaYuasa2022}. For classical systems, it is just the acquired information~\cite{Sagawa2012a}. For quantum systems, on the other hand, the backaction of quantum measurement reduces the acquired information, and the total information gain is given by $I_\mathrm{ms}=\chi_\mathrm{ms}-\Delta S_\mathrm{ms}$, as discussed in Refs.~\cite{SagawaUeda2008, FunoWatanabeUeda2013}. This quantity $I_\mathrm{ms}$ is called Groenewold information~\cite{Ozawa1986}.}. % 
This inequality~(\ref{eq:DeltaSfb}) shows that $\Delta S_\mathrm{fb}$ can be negative, i.e., the entropy can be reduced by feedback control.
The decrease in entropy $\Delta S_\mathrm{fb}<0$ not only enables demonic heat engines, but also allows the backaction of quantum measurements with $\Delta S_\mathrm{ms}>0$, so that energy can be injected, $E_\mathrm{ms}>0$, even in the steady cycle.

Thermal contact with a reservoir can also be used to reduce the entropy of the working substance.
According to the second law of thermodynamics,
\begin{equation}
\Delta S_\mathrm{th}\ge -\beta Q_\mathrm{out},
\label{eq:SecondLaw}
\end{equation}
where $\beta$ is the inverse temperature of the reservoir.
In quantum thermodynamics, a thermalization process is usually described by a CPTP map $\mathcal{E}_\mathrm{th}$ that preserves the Gibbs state at inverse temperature $\beta$.
The entropy change and the heat dissipated from the working substance to the thermal reservoir are defined as $\Delta S_\mathrm{th}=S\bm{(}\mathcal{E}_\mathrm{th}(\rho)\bm{)}-S(\rho)$ and $Q_\mathrm{out}=\Tr(H\rho)-\Tr\bm{(}H\mathcal{E}_\mathrm{th}(\rho)\bm{)}$, respectively (see, e.g., Ref.~\cite{Sagawa2022}).
The inequality~(\ref{eq:SecondLaw}) shows that the entropy reduction, $\Delta S_\mathrm{th}<0$, is possible by dissipating heat, $Q_\mathrm{out}>0$.
Such entropy reduction allows the entropy increase $\Delta S_\mathrm{ms}>0$ induced by quantum measurements even in the steady cycle, thereby activating the measurement backaction.

Finally, we comment on the thermodynamic interpretation of measurement backaction in quantum measurement-powered engines.
The inequality~(\ref{eq:UnitalEntropyGain}), together with the necessary and sufficient condition~(\ref{eq:MainTheorem}) for equality, implies that measurement backaction inevitably increases entropy~\cite{footnote:BareVSUnital}. %\footnote{Note that this holds for bare quantum measurements, but not for all unital quantum measurements. Consider, for instance, a unital (but not bare) quantum measurement $\mathcal{M}_\mathrm{unital}(\rho)=P_0\rho P_0+\sigma_x^{(12)}(\openone-P_0)\rho(\openone-P_0)\sigma_x^{(12)}$ for a three-level system, where $P_0=|0\rangle\langle0|$ and $\sigma_x^{(12)}=|1\rangle\langle 2|+|2\rangle\langle 1|$. This measurement $\mathcal{M}_\mathrm{unital}$ disturbs pure states $\rho=|1\rangle\langle 1|$ and $\rho=|2\rangle\langle2|$ while preserving the entropy.} %
The associated energy change has therefore been suggested to be interpreted as ``quantum heat''~\cite{ElouardHerrera-MartiClusel2017}.
However, unlike standard thermodynamic heat, measurement backaction never dissipates entropy from the working substance (see Table~\ref{table:ComparingEntropyGain}, where we summarize the entropy changes by various quantum thermodynamic operations).
This unidirectional entropy flow fundamentally distinguishes the energy change induced by measurement backaction from heat.
In particular, while work can be extracted from heat engines operating between reservoirs at different temperatures, an engine driven purely by bare quantum measurements cannot extract work in the steady regime, even when multiple measurements are applied in place of heat baths.
Measurement backaction cannot serve as a thermodynamic heat source in steady-cycle engines.

%++++++++++++++++++
\begin{table}[tb]
\caption{
Quantum thermodynamic operations.
}
\begin{ruledtabular}
\begin{tabular}{cc}
thermodynamic operations
&
entropy gain
\\
\hline
unitary operation
&
$\Delta S_U=0$
\\
thermalization (heat)
&
$\Delta S_\mathrm{th}\geq{-}\beta Q_\mathrm{out}$
\\
bare measurement (backaction)
&
$\Delta S_\mathrm{ms}\ge0$
\\
unitary feedback (information)
&
$\Delta S_\mathrm{fb}\geq{-}\chi_\mathrm{ms}$
\end{tabular}
\end{ruledtabular}
\label{table:ComparingEntropyGain}
\end{table}
%++++++++++++++++++

%%%%%%%%%%%%%%%%%%%%%%%%%%%%%%%%%%%%%%%%%%%%%%%%%%%%%%%%%%%%%%%%%%%%%%%%%%%%%%%
\section{Conclusion}
\label{section:Conclusion}
In this paper, we investigated a quantum heat engine powered purely by quantum measurements, i.e., by bare quantum measurements.
We proved a no-go result for work extraction from such an engine operating in the steady regime.
This implies the necessity of an entropy-decreasing process, such as feedback control or thermal contact, for work extraction in steady-cycle measurement-powered engines.
This result is based on~(\ref{eq:UnitalEntropyGain}) and~(\ref{eq:MainTheorem}), which imply that the backaction of a bare quantum measurement inevitably increases the entropy of the measured system.
In the steady regime, such an entropy increase is prohibited, and therefore bare quantum measurements become nondisturbing.
As a consequence, they do not inject energy into the working substance, and no work can be extracted from the engine.

Although our result is formulated for nonselective bare quantum measurements, its proof applies to a broader class of quantum operations. 
In fact, each bare measurement may be replaced by a CPTP map that is self-dual and has nonnegative eigenvalues. 
Theorem~\ref{thm:MainTheoremAppendix} in Appendix~\ref{appendix:ProofMainTheorem} guarantees that such a quantum operation is nondisturbing whenever it preserves the entropy of the working substance, and the same no-go result follows. 
For instance, a partial dephasing channel that leaves diagonal blocks unchanged and multiplies off-diagonal blocks by a nonnegative factor smaller than $1$ belongs to this class of quantum operations.
Such quantum operations can be regarded as instantaneous ``kicks'' rather than measurements.

Our result is general and applies to arbitrary quantum measurement-powered engines with finite-dimensional working substances.
Extending the result to infinite-dimensional systems is an important direction for future work (see, e.g., Refs.~\cite{ElouardJordan2018, YiTalknerKim2017, DingYiKim2018, JordanElouardAuffeves2020, ManikandanElouardMurch2022, JussiauBresqueAuffeves2023, Rodin2023}), since many physically relevant measurements, such as position and momentum measurements or photon detection, naturally involve infinite-dimensional systems.

In this work, we focused on the role of measurement backaction as an energetic resource in quantum heat engines.
Related studies have explored the use of measurement backaction to induce and control energy transport~\cite{BuffoniSolfanelliVerrucchi2019, BhandariJordan2022, YamamotoTokuraKato2022, YanJing2023}.
It would be interesting to develop a unified thermodynamic framework that exploits measurement backaction as a resource and is applicable to a variety of scenarios including energy transport phenomena.

%%%%%%%%%%%%%%%%%%%%%%%%%%%%%%%%%%%%%%%%%%%%%%%%%%%%%%%%%%%%%%%%%%%%%%%%%%%%%%%
\acknowledgments
We are grateful to Taiki Fujimoto for valuable discussions on the mathematical details underlying the main result.
This work was supported by JSPS KAKENHI Grants No.~JP24K22864 and No.~JP24K06904 from the Japan Society for the Promotion of Science (JSPS).

%%%%%%%%%%%%%%%%%%%%%%%%%%%%%%%%%%%%%%%%%%%%%%%%%%%%%%%%%%%%%%%%%%%%%%%%%%%%%%%
\appendix
%%%%%%%%%%%%%%%%%%%%%%%%%%%%%%%%%%%%%%%%%%%%%%%%%%%%%%%%%%%%%%%%%%%%%%%%%%%%%%%
\section{Proof of Nondisturbance Condition~(\ref{eq:MainTheorem})}
\label{appendix:ProofMainTheorem}
We prove the necessary and sufficient condition~(\ref{eq:MainTheorem}) for equality in the entropy inequality~(\ref{eq:UnitalEntropyGain}) under a nonselective bare quantum measurement $\mathcal{M}$. 
To this end, we prove the following theorem.
\begin{thm}
\label{thm:MainTheoremAppendix}
Let $\mathcal{M}$ be a positive and trace-preserving (PTP) map.
Assume that it is self-dual, $\mathcal{M}=\mathcal{M}^\dag$, and that its eigenvalues are nonnegative. 
Then, $\mathcal{M}$ preserves the von Neumann entropy $S(\sigma)=-\Tr(\sigma\log\sigma)$ of a positive-semidefinite operator $\sigma\geq0$ (in particular a density operator $\rho$) if and only if $\sigma$ is a fixed point of $\mathcal{M}$, i.e.,
\begin{equation}
S\bm{(}\mathcal{M}(\sigma)\bm{)}
=
S(\sigma)
\quad
\Leftrightarrow
\quad
\mathcal{M}(\sigma)=\sigma.
\label{eq:MainTheoremAppendix}
\end{equation}
\end{thm}
A nonselective bare quantum measurement $\mathcal{M}(\rho)=\sum_sM_s\rho M_s$ with positive-semidefinite measurement operators $M_s\ge0$ satisfying the Kraus condition $\sum_sM_s^2=\openone$ fulfills the conditions required for Theorem~\ref{thm:MainTheoremAppendix}\@.
To see it, let us use the Hilbert-Schmidt inner product $(X|Y)=\Tr(X^\dag Y)$ and introduce the matrix representation $\hat{\mathcal{M}}$ of $\mathcal{M}$ discussed in Refs.~\cite{Havel2003,GilchristTernoWood2011,Wood2015,Watrous2018,Hahn2022}.
The matrix $\hat{\mathcal{M}}$ acts on a vector $|X)$ as $\hat{\mathcal{M}}|X)=|\mathcal{M}(X))$.
First, since a nonselective bare quantum measurement $\mathcal{M}$ is CPTP, it is PTP\@.
Next, a nonselective bare quantum measurement $\mathcal{M}$ is self-dual, $\mathcal{M}=\mathcal{M}^\dag$.
Indeed,
\begingroup
\allowdisplaybreaks
\begin{align}
(\mathcal{M}^\dag(X)|Y)
&=
(X|\mathcal{M}(Y))
\nonumber\\
&=
\sum_s \Tr(X^\dagger M_s Y M_s)
\nonumber\\
&=
\sum_s \Tr(M_s X^\dagger M_s Y)
\nonumber\\
&=
(\mathcal{M}(X)|Y),\quad\forall X,Y,
\end{align}
\endgroup
which shows $\mathcal{M}=\mathcal{M}^\dag$.
This self-duality $\mathcal{M}=\mathcal{M}^\dag$ is reflected in the Hermitianity of the matrix $\hat{\mathcal{M}}$,
\begingroup
\allowdisplaybreaks
\begin{align}
(X|\hat{\mathcal{M}}^\dag|Y)
&=
(\mathcal{M}(X)|Y)
\nonumber\\
&=
(\mathcal{M}^\dag(X)|Y)
\nonumber\\
&=
(X|\mathcal{M}(Y))
\nonumber\\
&=
(X|\hat{\mathcal{M}}|Y),\quad\forall X,Y,
\label{eq:SelfDualHermitianity}
\end{align}
\endgroup
which shows $\hat{\mathcal{M}}=\hat{\mathcal{M}}^\dag$.
Furthermore, the eigenvalues of $\hat{\mathcal{M}}$ and hence of $\mathcal{M}$ are nonnegative, since
\begingroup
\allowdisplaybreaks
\begin{align}
(X|\hat{\mathcal{M}}|X)
&=
(X|\mathcal{M}(X))
\nonumber\\
&=
\sum_s \Tr(X^\dagger M_s X M_s)
\nonumber\\
&=
\sum_s \|\sqrt{M_s} X \sqrt{M_s}\|_\mathrm{HS}^2
\ge 0,\quad\forall X,
\label{eq:MhatNonnegative}
\end{align}
\endgroup
where $\|X\|_\mathrm{HS}=\sqrt{(X|X)}=\sqrt{\Tr(X^\dagger X)}$ is the Hilbert-Schmidt norm.
The positivity~(\ref{eq:MhatNonnegative}) actually implies the Hermitianity $\hat{\mathcal{M}}=\hat{\mathcal{M}}^\dag$ and hence the self-duality $\mathcal{M}=\mathcal{M}^\dag$.
Therefore, Theorem~\ref{thm:MainTheoremAppendix} applies for a nonselective bare quantum measurement $\mathcal{M}$, and the necessary and sufficient condition~(\ref{eq:MainTheorem}) for equality in the entropy inequality~(\ref{eq:UnitalEntropyGain}) follows.

We prove Theorem~\ref{thm:MainTheoremAppendix} using the following lemma.
\begin{lemma}[(1/2)-contractivity lemma]
\label{lem:1/2-contractivity}
Let $\mathcal{M}$ be a PTP map.
Assume that it is self-dual, $\mathcal{M}=\mathcal{M}^\dag$, and that its eigenvalues are nonnegative.
Then, for any operator $X$,
\begin{equation}
\|\mathcal{M}(X)\|_\mathrm{HS}^2
\le
(X|\mathcal{M}(X))
\le
\|X\|_\mathrm{HS}^2.
\label{eq:1/2-contractivity}
\end{equation}
Moreover, equality in the second inequality, $(X|\mathcal{M}(X))=\|X\|_\mathrm{HS}^2$, holds if and only if $X$ is a fixed point of $\mathcal{M}$, i.e., $\mathcal{M}(X)=X$.
\end{lemma}
A proof of this lemma is given in Appendix~\ref{appendix:1/2-contractivity}\@.
The saturation of the second inequality, $(X|\mathcal{M}(X))=\|X\|_\mathrm{HS}^2$, implies $\mathcal{M}(X)=X$, and further implies the saturation of the first inequality, $\|\mathcal{M}(X)\|_\mathrm{HS}^2=(X|\mathcal{M}(X))$.
However, the converse is not true: equality in the first inequality, $\|\mathcal{M}(X)\|_\mathrm{HS}^2=(X|\mathcal{M}(X))$, does not necessarily imply equality in the second inequality~\cite{footnote:Lemma}. %\footnote{Consider, for instance, a projective measurement $\mathcal{M}(X)=P_0X P_0+P_1X P_1$ for a two-level system, with $P_0=|0\rangle\langle0|$ and $P_1=|1\rangle\langle1|$, and apply it to $X=p_0|0\rangle\langle0|+p_1|1\rangle\langle1|+c|0\rangle\langle1|+c^*|1\rangle\langle0|$. We have $\mathcal{M}^2(X)=\mathcal{M}(X)=p_0|0\rangle\langle0|+p_1|1\rangle\langle1|\neq X$, while $\|\mathcal{M}(X)\|_\mathrm{HS}^2=(X|\mathcal{M}^2(X))=(X|\mathcal{M}(X))$. Equality in the first inequality in~(\ref{eq:1/2-contractivity}) holds, while $X$ is not a fixed point of $\mathcal{M}$ and equality in the second inequality in~(\ref{eq:1/2-contractivity}) does not hold.}. %

A similar lemma is proved in Ref.~\cite{Wen2026}, but under a stronger condition of complete positivity.
In addition, our lemma includes intermediate inequalities for $(X|\mathcal{M}(X))$ in~(\ref{eq:1/2-contractivity}).
Note that the Hilbert-Schmidt norm (as well as other Schatten norms) is contractive, $\|\Phi(X)\|_\mathrm{HS}\leq\|X\|_\mathrm{HS}$, under any PTP and unital map $\Phi$~\cite{Perez-GarciaWolfPetz2006}. 
It is also the case for a PTP and self-dual map $\mathcal{M}$, since such a map $\mathcal{M}$ is also unital.
The intermediate inequalities in~(\ref{eq:1/2-contractivity}) look like the contractivity under the ``square-root'' of $\mathcal{M}$, and we call our lemma ``(1/2)-contractivity lemma.''

Let us prove Theorem~\ref{thm:MainTheoremAppendix} using Lemma~\ref{lem:1/2-contractivity}\@.
\begin{proof}[Proof of Theorem~\ref{thm:MainTheoremAppendix}]
The implication $\mathcal{M}(\sigma)=\sigma \Rightarrow S\bm{(}\mathcal{M}(\sigma)\bm{)}=S(\sigma)$ is trivial. 
We prove the converse.

Assume that $S\bm{(}\mathcal{M}(\sigma)\bm{)}=S(\sigma)$. 
Since $\mathcal{M}$ is PTP and self-dual, it is unital.
Thus, $\sigma$ majorizes $\mathcal{M}(\sigma)$, and the inequality $S\bm{(}\mathcal{M}(\sigma)\bm{)}\ge S(\sigma)$ generally holds~\cite{LiBusch2013}.
Equality in this entropy inequality then implies equality in the majorization inequalities, which implies that $\sigma$ and $\mathcal{M}(\sigma)$ have the same spectrum, as shown in Ref.~\cite{LiBusch2013}. 
Hence, there exists a unitary operator $U$ such that
\begin{equation}
\mathcal{M}(\sigma)=U\sigma U^\dagger.
\end{equation}
Therefore,
\begin{equation}
\|\mathcal{M}(\sigma)\|_\mathrm{HS}^2
=
\|U\sigma U^\dagger\|_\mathrm{HS}^2
=
\|\sigma\|_\mathrm{HS}^2.
\end{equation}
Using this in Lemma~\ref{lem:1/2-contractivity}, we obtain
\begin{equation}
(\sigma|\mathcal{M}(\sigma))
=
\|\sigma\|_\mathrm{HS}^2.
\end{equation}
Thus, equality holds in the second inequality in~(\ref{eq:1/2-contractivity}), which implies $\mathcal{M}(\sigma)=\sigma$. 
\end{proof}

An alternative proof of Theorem~\ref{thm:MainTheoremAppendix} is available, based on the entropy-gain inequality proved in Ref.~\cite{BuscemiDasWilde2016}.
See Appendix~\ref{app:AlternativeProof}\@.

%%%%%%%%%%%%%%%%%%%%%%%%%%%%%%%%%%%%%%%%%%%%%%%%%%%%%%%%%%%%%%%%%%%%%%%%%%%%%%%
\section{Proof of (1/2)-Contractivity Lemma}
\label{appendix:1/2-contractivity}
We prove Lemma~\ref{lem:1/2-contractivity}\@.
The self-duality of $\mathcal{M}$ implies the Hermitianity of $\hat{\mathcal{M}}$, as shown in~(\ref{eq:SelfDualHermitianity}).
The Hermitian matrix $\hat{\mathcal{M}}$ thus admits the spectral decomposition
\begin{equation}
\hat{\mathcal{M}}
=
\sum_\ell \lambda_\ell |\lambda_\ell)(\lambda_\ell|,
\label{eq:SpecDecomp}
\end{equation}
with real eigenvalues $\{\lambda_\ell\}$ and a complete set of orthonormal basis vectors $\{|\lambda_\ell)\}$ satisfying the orthonormality $(\lambda_\ell|\lambda_{\ell'})=\delta_{\ell\ell'}$ and the completeness $\sum_\ell|\lambda_\ell)(\lambda_\ell|=\hat{\mathcal{I}}$, where $\hat{\mathcal{I}}$ is the matrix representation of the identity map.
The real eigenvalues $\{\lambda_\ell\}$ are confined in $\lambda_\ell\in[0,1]$, since the eigenvalues of a PTP map $\mathcal{M}$ in general are confined within the unit disk on the complex plane~\cite{Wolf2012, BurgarthFacchiNakazato2020, AmatoFacchiKonderak2023, Hahn2025}, and the eigenvalues of $\mathcal{M}$ are assumed to be nonnegative in Lemma~\ref{lem:1/2-contractivity}\@.

We now prove the inequalities in~(\ref{eq:1/2-contractivity}). 
Using the spectral decomposition~(\ref{eq:SpecDecomp}),
\begin{align}
\|\mathcal{M}(X)\|_\mathrm{HS}^2
={}&
(\mathcal{M}(X)| \mathcal{M}(X))
\nonumber\\
={}&
(X|\hat{\mathcal{M}}^2|X)
\nonumber\\
={}&
\sum_\ell\lambda_\ell^2|(\lambda_\ell|X)|^2
\nonumber\\
\leq{}&
\sum_\ell\lambda_\ell|(\lambda_\ell|X)|^2
\nonumber\\
={}&
(X|\hat{\mathcal{M}}|X)
\nonumber\\
={}&
(X|\mathcal{M}(X)),
\end{align}
which proves the first inequality. 
By further bounding it as
\begin{align}
(X|\mathcal{M}(X))
={}&
(X|\hat{\mathcal{M}}|X)
\nonumber\\
={}&
\sum_\ell\lambda_\ell|(\lambda_\ell|X)|^2
\nonumber\\
\leq{}&
\sum_\ell|(\lambda_\ell|X)|^2
\nonumber\\
={}&
(X|X)
\nonumber\\
={}&
\|X\|_\mathrm{HS}^2,
\end{align}
which proves the second inequality.

The second inequality in~(\ref{eq:1/2-contractivity}) can also be obtained in the following way. 
Using the self-duality $\mathcal{M}=\mathcal{M}^\dag$, we have
\begingroup
\allowdisplaybreaks
\begin{align}
&\|\mathcal{M}(X)-X\|_\mathrm{HS}^2
\nonumber\\
&\qquad
=
(\mathcal{M}(X)-X|\mathcal{M}(X)-X)
\nonumber\\
&\qquad
=
(\mathcal{M}(X)|\mathcal{M}(X))
-(\mathcal{M}(X)|X)
\nonumber\\
&\qquad\hphantom{{}=(\mathcal{M}(X)|\mathcal{M}(X))
}
-(X|\mathcal{M}(X))
+(X|X)
\nonumber\\
&\qquad
=
\|\mathcal M(X)\|_\mathrm{HS}^2
-2(X|\mathcal{M}(X))
+\|X\|_\mathrm{HS}^2
.
\end{align}
\endgroup
This can be bounded, using the first inequality in~(\ref{eq:1/2-contractivity}), by
\begin{equation}
0\leq
\|\mathcal M(X)-X\|_\mathrm{HS}^2
\leq
\|X\|_\mathrm{HS}^2-(X|\mathcal{M}(X))
,
\end{equation}
which is nothing but the second inequality in~(\ref{eq:1/2-contractivity}). 
The equality $\|X\|_\mathrm{HS}^2=(X|\mathcal{M}(X))$ holds if and only if $\|\mathcal M(X)-X\|_\mathrm{HS}^2=0$, that is, if and only if $\mathcal M(X)=X$.

If $\mathcal{M}$ is not only positive but also completely positive, which is the case for a nonselective bare quantum measurement, the second inequality in~(\ref{eq:1/2-contractivity}) can be proven without relying on the first inequality in~(\ref{eq:1/2-contractivity}) or on the nonnegativity of the eigenvalues of $\mathcal{M}$. 
Note that a self-dual CPTP map $\mathcal{M}$ admits a Kraus representation $\mathcal M(X)=\sum_sA_sXA_s$ with Hermitian Kraus operators $A_s=A_s^\dag$~\cite{Wolf2012,O'MearaPereira2013}.
Using this with the Kraus condition $\sum_sA_s^2=\openone$, we obtain 
\begin{align}
&\|X\|_\mathrm{HS}^2-(X|\mathcal{M}(X))
\nonumber\\
&\quad
=
\frac{1}{2}
\sum_s\Bigl(
\Tr[A_s^2(X^\dagger X+XX^\dag)]
-2\Tr(X^\dagger A_sX A_s)
\Bigr)
\nonumber\\
&\quad
=
\frac{1}{2}
\sum_s
\Tr(
A_sX^\dagger XA_s+X^\dag A_sA_sX
\nonumber\\[-3truemm]
&\quad
\hphantom{{}=\frac{1}{2}\sum_s\Tr(}
{}-X^\dagger A_sX A_s
-A_sX^\dagger A_sX
)
\nonumber\\
&\quad
=
\frac{1}{2}
\sum_s
\|
[X,A_s]
\|_\mathrm{HS}^2
\nonumber\\
&\quad\ge0.
\vphantom{\frac{1}{2}}
\label{eq:ProofOfLemma2}
\end{align}
The equality $\|X\|_\mathrm{HS}^2=(X|\mathcal{M}(X))$ holds if and only if $[X,A_s]=0$ for all $s$, which implies that
\begin{equation}
\mathcal{M}(X)=\sum_sA_sX A_s=\sum_sA_s^2X=X,
\end{equation}
i.e., $X$ is a fixed point of $\mathcal{M}$.

\section{Alternative Proof of Theorem~\ref{thm:MainTheoremAppendix}}
\label{app:AlternativeProof}
Here, we present an alternative proof of Theorem~\ref{thm:MainTheoremAppendix}\@.
In Ref.~\cite{BuscemiDasWilde2016}, the following entropy-gain inequality is proved, which we restate here as a proposition.
\begin{prop}[Theorem~1 of Ref.~\cite{BuscemiDasWilde2016}]
\label{prop:BuscemiInequality}
Let $\rho$ be a density operator and let $\mathcal{N}$ be a PTP map.
Then, 
\begin{equation}
S\bm{(}\mathcal{N}(\rho)\bm{)}
-
S(\rho)
\ge D\bm{(}\rho\|(\mathcal{N}^\dag\circ\mathcal{N})(\rho)\bm{)},
\label{eqn:BuscemiInequality}
\end{equation}
where $D(\sigma\|\sigma')=\Tr[\sigma(\log\sigma-\log\sigma')]$ is the quantum relative entropy for positive-semidefinite operators $\sigma$ and $\sigma'$.
\end{prop}
The inequality~(\ref{eqn:BuscemiInequality}) holds also for an arbitrary positive-semidefinite operator $\sigma$ in place of the density operator $\rho$.
If $\mathcal{N}$ is additionally subunital, $\mathcal{N}(\openone)\le\openone$, then $\mathcal{N}^\dag$ is trace-nonincreasing, and~\cite{BuscemiDasWilde2016}
\begin{equation}
D\bm{(}\rho\|(\mathcal{N}^\dag\circ\mathcal{N})(\rho)\bm{)}\ge0.
\end{equation}

We can apply this proposition to prove Theorem~\ref{thm:MainTheoremAppendix}, since $\mathcal{M}$ of Theorem~\ref{thm:MainTheoremAppendix} is also PTP\@.
In Theorem~\ref{thm:MainTheoremAppendix}, it is further assumed to be self-dual, $\mathcal{M}=\mathcal{M}^\dag$, with nonnegative eigenvalues, $\lambda_\ell\ge0$.
\begin{proof}[Alternative proof of Theorem~\ref{thm:MainTheoremAppendix}]
The implication $\mathcal{M}(\sigma)=\sigma \Rightarrow S\bm{(}\mathcal{M}(\sigma)\bm{)}=S(\sigma)$ is trivial.
We prove the converse.

Assume that $S\bm{(}\mathcal{M}(\sigma)\bm{)}=S(\sigma)$.
Then, the inequality~(\ref{eqn:BuscemiInequality}) implies $D\bm{(}\sigma\|(\mathcal{M}^\dag\circ\mathcal{M})(\sigma)\bm{)}\le0$.
Since $\mathcal{M}$ is TP and self-dual, it is unital, $\mathcal{M}(\openone)=\openone$, and hence, $D\bm{(}\sigma\|(\mathcal{M}^\dag\circ\mathcal{M})(\sigma)\bm{)}\ge0$.
Consequently, $D\bm{(}\sigma\|(\mathcal{M}^\dag\circ\mathcal{M})(\sigma)\bm{)}=0$.
Since $\mathcal{M}=\mathcal{M}^\dag$, the equality for the quantum relative entropy yields
\begin{equation}
(\mathcal{M}^\dag\circ\mathcal{M})(\sigma)=\mathcal{M}^2(\sigma)=\sigma.
\end{equation}
Using the spectral decomposition~(\ref{eq:SpecDecomp}) of $\mathcal{M}$, this condition reads
\begin{equation}
\mathcal{M}^2(\sigma)
=\sum_\ell\lambda_\ell^2|\lambda_\ell)(\lambda_\ell|\sigma)
=\sum_\ell|\lambda_\ell)(\lambda_\ell|\sigma)
=\sigma.
\end{equation}
This implies $\lambda_\ell^2=1$ whenever $(\lambda_\ell|\sigma)\neq0$.
Since the eigenvalues $\{\lambda_\ell\}$ of $\mathcal{M}$ are assumed to be nonnegative, $\lambda_\ell\ge0$, we have $\lambda_\ell=1$ whenever $(\lambda_\ell|\sigma)\neq0$.
Therefore,
\begin{equation}
\mathcal{M}(\sigma)
=\sum_\ell\lambda_\ell|\lambda_\ell)(\lambda_\ell|\sigma)
=\sum_\ell|\lambda_\ell)(\lambda_\ell|\sigma)
=\sigma.
\end{equation}
\end{proof}

Note that the equivalence
\begin{equation}
S\bm{(}\Phi(\sigma)\bm{)}=S(\sigma)
\quad\Leftrightarrow\quad
(\Phi^\dag\circ\Phi)(\sigma)=\sigma
\label{eqn:ZhangEquivalence}
\end{equation}
is proved for unital CPTP maps $\Phi$ in Ref.~\cite{ZhangWu2011}.
Proposition~\ref{prop:BuscemiInequality}, together with the monotonicity of the entropy under unital PTP maps, extends this equivalence to unital PTP maps.
\begin{corol}
\label{corol:GeneralizedZhangEquivalence}
Let $\sigma$ be a positive-semidefinite operator and let $\Phi$ be a unital PTP map.
Then, the equivalence~(\ref{eqn:ZhangEquivalence}) holds.
\end{corol}
\begin{proof}
Since $\Phi$ is a unital PTP map, so is $\Phi^\dag$.
Therefore,
\begin{equation}
S(\sigma)
\leq
S\bm{(}\Phi(\sigma)\bm{)}
\leq
S\bm{(}(\Phi^\dag\circ\Phi)(\sigma)\bm{)}.
\label{eqn:InequalityChain}
\end{equation}
The forward implication in~(\ref{eqn:ZhangEquivalence}) follows from Proposition~\ref{prop:BuscemiInequality} and the nonnegativity of the quantum relative entropy.
The converse follows immediately from the chain of inequalities in~(\ref{eqn:InequalityChain}).
\end{proof}

%%%%%%%%%%%%%%%%%%%%%%%%%%%%%%%%%%%%%%%%%%%%%%%%%%%%%%%%%%%%%%%%%%%%%%%%%%%%%%%
%\bibliographystyle{apsrev4-1}
%\bibliographystyle{prsty-title-hyperref}
%\bibliography{MeasurementEngine}

\begin{thebibliography}{10}
\bibitem{footnote:BareVSUnital}
Note that this holds for bare quantum measurements, but not for all unital quantum measurements. Consider, for instance, a unital (but not bare) quantum measurement $\mathcal{M}_\mathrm{unital}(\rho)=P_0\rho P_0+\sigma_x^{(12)}(\openone-P_0)\rho(\openone-P_0)\sigma_x^{(12)}$ for a three-level system, where $P_0=|0\rangle\langle0|$ and $\sigma_x^{(12)}=|1\rangle\langle 2|+|2\rangle\langle 1|$. This measurement $\mathcal{M}_\mathrm{unital}$ disturbs pure states $\rho=|1\rangle\langle 1|$ and $\rho=|2\rangle\langle2|$ while preserving the entropy.
\bibitem{footnote:Lemma}
Consider, for instance, a projective measurement $\mathcal{M}(X)=P_0X P_0+P_1X P_1$ for a two-level system, with $P_0=|0\rangle\langle0|$ and $P_1=|1\rangle\langle1|$, and apply it to $X=p_0|0\rangle\langle0|+p_1|1\rangle\langle1|+c|0\rangle\langle1|+c^*|1\rangle\langle0|$. We have $\mathcal{M}^2(X)=\mathcal{M}(X)=p_0|0\rangle\langle0|+p_1|1\rangle\langle1|\neq X$, while $\|\mathcal{M}(X)\|_\mathrm{HS}^2=(X|\mathcal{M}^2(X))=(X|\mathcal{M}(X))$. The equality in the first inequality in~(\ref{eq:1/2-contractivity}) holds, while $X$ is not a fixed point of $\mathcal{M}$ and the equality in the second inequality in~(\ref{eq:1/2-contractivity}) does not hold.
\end{thebibliography}
%\end{document}

\end{document}